\documentclass[conference]{IEEEtran}
\IEEEoverridecommandlockouts

\usepackage{cite}
\usepackage{amsmath,amssymb,amsfonts}
\usepackage{algorithmic}
\usepackage{graphicx}
\usepackage{textcomp}
\usepackage{xcolor}

\usepackage[caption=false,font=footnotesize]{subfig}

\usepackage{booktabs}

\usepackage[section]{placeins}

\def\BibTeX{{\rm B\kern-.05em{\sc i\kern-.025em b}\kern-.08em
    T\kern-.1667em\lower.7ex\hbox{E}\kern-.125emX}}

\begin{document}

\title{Radar-Based Fall Detection for Assisted Living: A Digital-Twin Representation Case Study}

\author{
\IEEEauthorblockN{
Sebastian Ratto V\IEEEauthorrefmark{1},
Huy Trinh\IEEEauthorrefmark{1},
Ahmed N. Sayed\IEEEauthorrefmark{1},
Abdelrahman Elbadrawy\IEEEauthorrefmark{1},
Arien Sligar\IEEEauthorrefmark{2},
and George Shaker\IEEEauthorrefmark{1}
}
\IEEEauthorblockA{\IEEEauthorrefmark{1}University of Waterloo, Canada}
\IEEEauthorblockA{\IEEEauthorrefmark{2}Synopsys Inc., USA}
}

\maketitle

\begin{abstract}
Obtaining data on high-impact falls from older adults is ethically difficult, yet these rare events cause many fall-related health problems. As a result, most radar-based fall detectors are trained on staged falls from young volunteers, and representation choices are rarely tested against the radar signals from dangerous falls. This paper uses a frequency-modulated continuous-wave (FMCW) radar digital twin as a single simulated room testbed to study how representation choice affects fall/non-fall discrimination. From the same simulated range--Doppler sequence, Doppler--time spectrograms, three-channel per-receiver spectrogram stacks, and time-pooled range--Doppler maps (RDMs) are derived and fed to an identical compact CNN under matched training on a balanced fall/non-fall dataset. In this twin, temporal spectrograms reach $98$--$99\%$ test accuracy with similar precision and recall for both classes, while static RDMs reach $89.4\%$ and show more variable training despite using the same backbone. A qualitative comparison between synthetic and measured fall spectrograms suggests that the twin captures gross Doppler--time structure, but amplitude histograms reveal differences in the distributions of amplitude values consistent with receiver processing not modeled in the twin. Because the twin omits noise and hardware impairments and is only qualitatively compared to a single measured example, these results provide representation-level guidance under controlled synthetic conditions rather than ready-to-use clinical performance in real settings.
\end{abstract}


\section{Introduction}
\label{sec:intro}

Falls are a leading cause of injury, functional decline, and mortality in older adults~\cite{who_falls_2021,cdc_falls_facts_2024}. Long lie times after unwitnessed falls increase complications even when the initial impact is mild. Wearable devices can detect impacts but suffer from poor adherence, while camera-based systems raise privacy concerns and degrade in low light. Millimeter-wave frequency-modulated continuous-wave (FMCW) radar has therefore become a candidate sensing modality for ambient assisted living because it operates in the dark, preserves visual privacy, and does not require user-worn hardware~\cite{Shaker2025MLReview,newaz2023_fall_detection_review}. High-impact falls from frail older adults are difficult and unsafe to stage~\cite{9305931}. As a result, many radar-based fall detectors are trained and tested on scripted falls from young volunteers or mannequins~\cite{Khan_2017,9305931,rs15082101}. These scenarios cover only part of real fall motion patterns~\cite{s24020648} and under-represent slow collapses, fainting-like events, and high-energy slips. Real-world falls are also rare, so even large deployments collect few positive examples and provide limited information about failure modes. Under these constraints, the choice of radar representation becomes particularly important. Designers must pick between time--frequency spectrograms and range--Doppler maps (among other options), typically under tight memory budgets and with little evidence about which representations highlight the motion patterns of clinically important falls. A model can score well on staged falls while relying on artifacts of a specific setup rather than on robust temporal or range--Doppler structure.

Electromagnetic digital twins (DTs) offer a way to probe such design choices before large measurement campaigns. By combining human motion libraries, 3D room geometries, and radar front-end models, DTs generate synthetic radar measurements under controlled, reproducible conditions~\cite{10.1007/978-3-031-97772-5_3,Chipengo2021HighFid,Ratto2025ITC,11052547, ahmedDrones2024, Antemratto}. A radar DT can simulate fall and non-fall trajectories at multiple positions and yaw angles without exposing human subjects to risk, and can be used as a testbed for examining how different representations behave when motion patterns are varied systematically.
% This paper uses a single-room $60~\text{GHz}$ FMCW radar DT as a controlled testbed to compare how different representations support binary fall versus non-fall discrimination. The goal is not to propose a new fall detector, but to fix a compact convolutional neural network (CNN) probe and use it as a standard lens for comparing inputs derived from the same synthetic range--Doppler data. All models share the same CNN backbone, preprocessing pipeline, and stratified train/test split, so differences in performance are more readily interpreted as effects of representation choice rather than architecture changes.

Within this framework, this paper makes three contributions. \textbf{(1)} It generates a balanced synthetic fall/non-fall dataset in a furnished assisted-living-style room using a Perceive-EM-based workflow and a ceiling-mounted $60~\text{GHz}$ FMCW radar model under a fixed stratified train/test split. \textbf{(2)} It performs a controlled comparison of three input modes derived from the same range--Doppler sequence: an all-receiver spectrogram (\texttt{spec}), a per-receiver spectrogram stack (\texttt{spec3}), and a time-pooled range--Doppler map (\texttt{rdm}), all using identical compact CNN probes and training settings. \textbf{(3)} It finds that, in this setting, temporal spectrogram modes achieve higher test accuracy and more balanced precision/recall than a static RDM mode, which tends to yield lower non-fall recall and more variable performance across epochs. It also provides a qualitative comparison between synthetic and measured fall spectrograms as a sanity check on gross signal structure, while documenting the limitations of this idealized DT and illustrating a difference in amplitude distributions for one representative fall.

% All results come from a single synthetic room and a single virtual radar with simplified channel and noise modeling. They should be interpreted as representation-level guidance under controlled synthetic conditions, not as clinical performance claims.

\section{Methodology}
\label{sec:method}

This section describes how the synthetic fall/non-fall dataset is constructed, how the radar digital twin generates measurements and standard FMCW products, and how these measurements are converted into input representations for a fixed CNN probe.

\begin{figure}[t]
  \centering
  \includegraphics[width=\columnwidth]{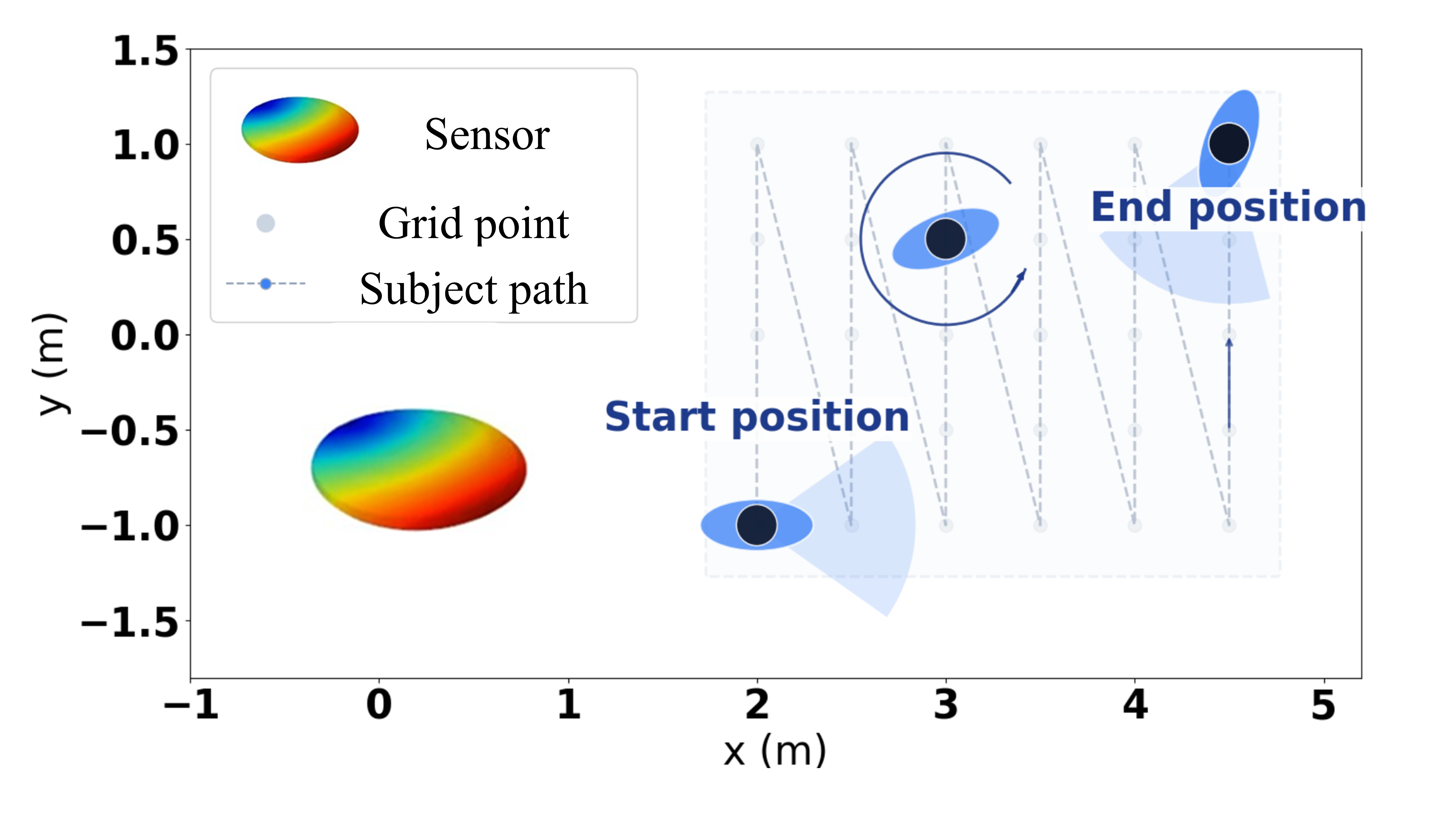}
  \caption{Discrete grid over subject position and yaw relative to the radar, with one example fall trajectory shown.}
  \label{fig:room_schematic}
\end{figure}

\begin{figure*}[t]
  \centering
  \includegraphics[width=0.9\textwidth]{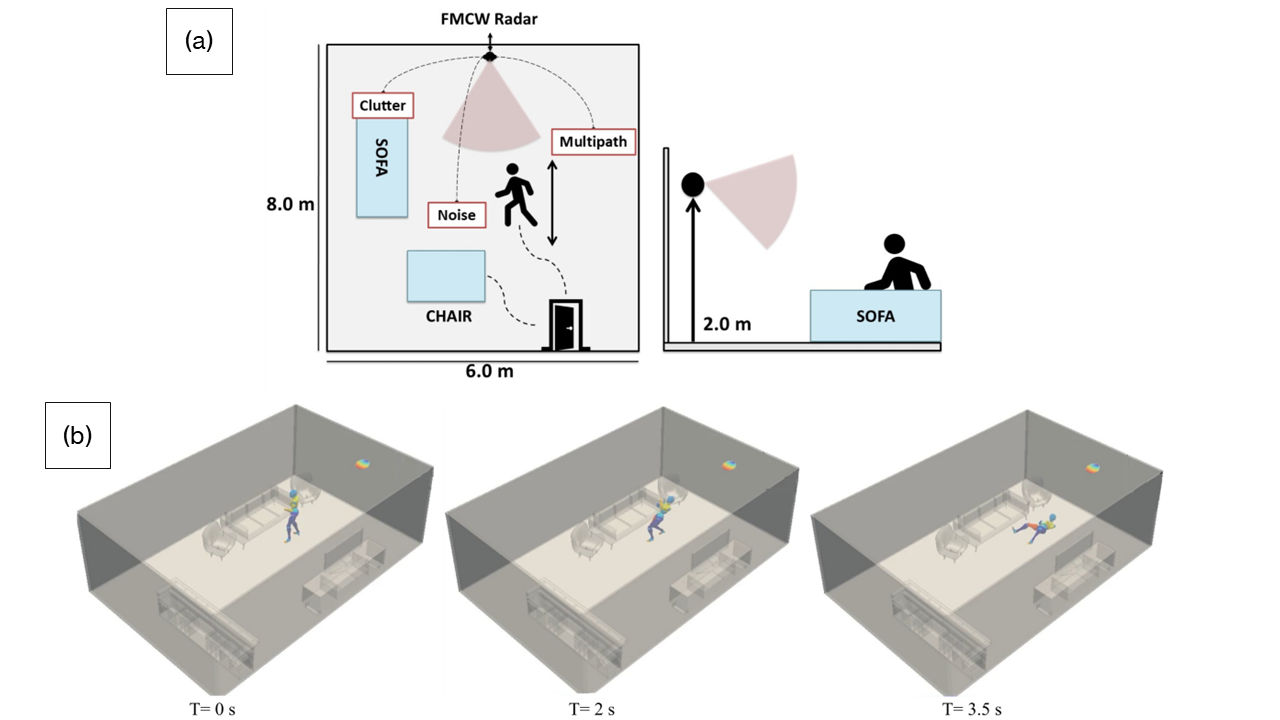}
  \caption{Digital-twin simulation setup. (a) Top-down and side-view schematic of the $6 \times 8~\text{m}$ furnished room, showing the ceiling-mounted FMCW radar, its field of view, and example reflecting objects in the modeled environment. (b) Perceive-EM renderings of one fall at three time instants.}
  \label{fig:dt_room}
\end{figure*}

\subsection{Dataset Construction}

The motion library covers a range of fall mechanics and everyday non-fall activities that are plausible in an assisted-living room while keeping class labels simple (fall versus non-fall). Human motions are built from animated meshes derived from the Mixamo database~\cite{mixamo}, a publicly available character animation library. Falls include knee-first, hip-first, forward, backward, side, stumble/trip, running, and hit-and-fall motions. Non-fall motions include walking, jumping in place, crouching to pick up an object, crouching then standing, sitting and clapping, and a baseball swing. Each motion template is assigned a binary label (fall/non-fall) and a numeric identifier; no severity labels are used in this study.

To probe how representations behave under changing subject pose and orientation, each motion is simulated on a discrete position--yaw grid relative to the radar. The grid spans five range positions, three cross-range positions, and sixteen yaw angles, yielding up to $5 \times 3 \times 16 = 240$ distinct subject pose configurations (unique position--yaw combinations) per motion. Simulations are launched programmatically via a batch script and Python driver that iterate over motion ID, position, and yaw, so the full grid can be regenerated deterministically. The position--yaw grid and an example trajectory are illustrated in Fig.~\ref{fig:room_schematic}.

After simulation, all clips are indexed by scanning the digital-twin output directory and recording, for each motion and pose configuration, the available radar files: all-receiver range--Doppler, all-receiver spectrogram, and three per-receiver spectrograms. A metadata table records the label (fall/non-fall), motion name, file paths, and pose configuration. Across all motions and poses this yields 6{,}959 indexed clips, with 5{,}519 labeled as falls and 1{,}440 as non-falls.

The raw dataset is strongly imbalanced toward falls. To avoid letting the classifier exploit this imbalance, all 1{,}440 non-fall clips are retained and 1{,}440 fall clips are drawn uniformly at random with a fixed seed, producing a balanced set of 2{,}880 clips. A stratified 80/20 split with respect to the fall/non-fall label then yields $N_{\text{train}} = 2{,}304$ training clips and $N_{\text{test}} = 576$ test clips, with equal numbers of falls and non-falls in each set. This split is generated once and reused for all representation modes so that only the input representation changes between experiments.

\begin{table}[t]
  \centering
  \caption{Simulated fall and non-fall motions in the digital-twin dataset.}
  \label{tab:motions}
  \small
  \setlength{\tabcolsep}{4pt}
  \renewcommand{\arraystretch}{0.9}
  \begin{tabular}{@{}clcc@{}}
    \toprule
    ID & Motion description                         & Class     & \# poses \\
    \midrule
     1 & Knee fall forward                          & Fall      &  64 \\
     2 & Hip fall with tuck 1 (backwards)           & Fall      &  62 \\
     4 & Full-body jump fall backwards              & Fall      &  65 \\
     6 & Full-body walk trip forward                & Fall      &  57 \\
     8 & Full-body running fall forward             & Fall      &  67 \\
     9 & Full-body get hit fall backward            & Fall      &  69 \\
    10 & Full-body backwards                        & Fall      &  56 \\
    11 & Hip fall landing on buttocks               & Fall      &  70 \\
    12 & Full-body forward                          & Fall      &  54 \\
    13 & Trip forward with flip                     & Fall      &  60 \\
    14 & Full-body backwards                        & Fall      &  64 \\
    15 & Hip fall with tuck 2 (backwards)           & Fall      &  64 \\
    16 & Hip fall with tuck 3 (backwards)           & Fall      &  65 \\
    17 & Forward knees-first fall                   & Fall      &  65 \\
    18 & Full-body sideways fall                    & Fall      &  55 \\
    19 & Backward fall (full body + hip-first)      & Fall      &  57 \\
    21 & Full-body side fall                        & Fall      &  59 \\
    22 & Diagonal forward knee fall                 & Fall      &  63 \\
    23 & Jump fall backwards                        & Fall      &  68 \\
    24 & Full-body backwards                        & Fall      &  66 \\
    25 & Hip fall backwards with tuck               & Fall      &  67 \\
    26 & Full-body stumble backwards                & Fall      &  54 \\
    27 & Knee fall forward                          & Fall      &  69 \\
    28 & Jump up                                    & Non-fall  & 240 \\
    29 & Crouch to pick up object                   & Non-fall  & 240 \\
    30 & Sitting clap                               & Non-fall  & 240 \\
    31 & Crouching then standing                    & Non-fall  & 240 \\
    32 & Baseball swing                             & Non-fall  & 240 \\
    33 & Walking forward                            & Non-fall  & 240 \\
    \bottomrule
  \end{tabular}
\end{table}

\subsection{Digital Twin and Radar Processing}

The digital twin is used to control geometry and propagation physics while removing measurement noise and subject recruitment constraints. Perceive-EM~\cite{perceive-em}, a commercial electromagnetic simulation tool for radar digital twins, uses a shooting-and-bouncing rays (SBR) solver to model indoor multipath propagation in furnished rooms~\cite{Chipengo2021HighFid,11052547}. Human meshes from Mixamo~\cite{mixamo} and room geometry are imported as triangle meshes with assigned $60~\text{GHz}$ material properties. Radar pose, antenna patterns, and FMCW waveform parameters are scripted via a Python API so that the same configuration can be reused across runs. Fig.~\ref{fig:room_schematic} and Fig.~\ref{fig:dt_room} summarize the room layout, position--yaw grid, and 3D digital-twin view.

\subsubsection{Radar configuration}

The virtual sensor is a ceiling-mounted $60~\text{GHz}$ FMCW radar with one transmit and three receive channels, configured via an Infineon-style JSON description. The active waveform (mode~1) uses a center frequency of $f_{c} = 60~\text{GHz}$ and an FMCW bandwidth of $B = 499.7~\text{MHz}$, with $N_{\text{s}} = 32$ analog-to-digital converter (ADC) samples per chirp at a $1~\text{mega-sample/s}~(1~\text{MS/s})$ sampling rate. Each coherent processing interval (CPI) contains $N_{\text{p}} = 64$ chirps over a duration of $T_{\text{CPI}} = 26.3~\text{ms}$, so the chirp repetition interval is $T_{\text{chirp}} = T_{\text{CPI}}/N_{\text{p}} \approx 0.41~\text{ms}$. The ideal range resolution is
\[
  \Delta R = \frac{c}{2B} \approx 0.30~\text{m},
\]
and the post-processing stage uses $N_{R} = 64$ range bins, covering approximately $0$--$19~\text{m}$. Doppler processing uses $N_{D} = 256$ FFT bins per CPI. With $N_{\text{p}} = 64$ chirps and wavelength $\lambda = c/f_{c} = 5~\text{mm}$, the theoretical Doppler resolution and unambiguous velocity are
\[
  \Delta v \approx \frac{\lambda}{2 T_{\text{CPI}}} \approx 0.10~\text{m/s},
  \qquad
  |v_{\max}| \approx \frac{\lambda}{4 T_{\text{chirp}}} \approx 3.0~\text{m/s},
\]
and the $256$-point Doppler FFT samples this interval. The antenna front-end consists of one vertically polarized parametric transmit patch (Tx1) and three vertically polarized parametric receive patches (Rx1--Rx3), each modeled with half-power beamwidths of $30.5^{\circ}$ in azimuth and $60.5^{\circ}$ in elevation and positioned within a few millimeters of the radar platform origin, forming a compact $1\times3$ array. The transmit channel is excited with $1~\text{W}$ input power; absolute power scaling is later removed by per-sample zero-mean, unit-variance (z-score) normalization of the radar representations, so the study focuses on relative spatial and temporal structure rather than link-budget limits.

For each simulated clip, the solver traces rays from the transmit aperture through the furnished scene, accumulates field contributions at the receive aperture, and exports complex baseband samples per antenna and per chirp. The receiver chain is idealized: no explicit thermal noise, oscillator phase noise, automatic gain control, or non-stationary clutter suppression stages are modeled, and the analog-to-digital converter is assumed linear over the dynamic range of interest. Under these modeling assumptions, the electromagnetic solution is intended to be the dominant source of scene-dependent variability, so differences between clips are driven by motion and pose rather than hardware artifacts.

Standard FMCW processing is applied offline. For each receive channel, a sequence of range--Doppler maps $X_t(r,q)$ is formed by applying a 2-D FFT over fast time and chirps for each coherent processing interval (CPI), with $t$ indexing the CPIs. Velocity--time spectrograms are then constructed by sampling $X_t(r,q)$ at a fixed range bin $r^\star \approx 3.9~\text{m}$ and stacking the resulting Doppler slices over time, converting magnitudes to decibels before resizing and normalization.

\begin{figure}[t]
  \centering
  \includegraphics[width=\columnwidth]{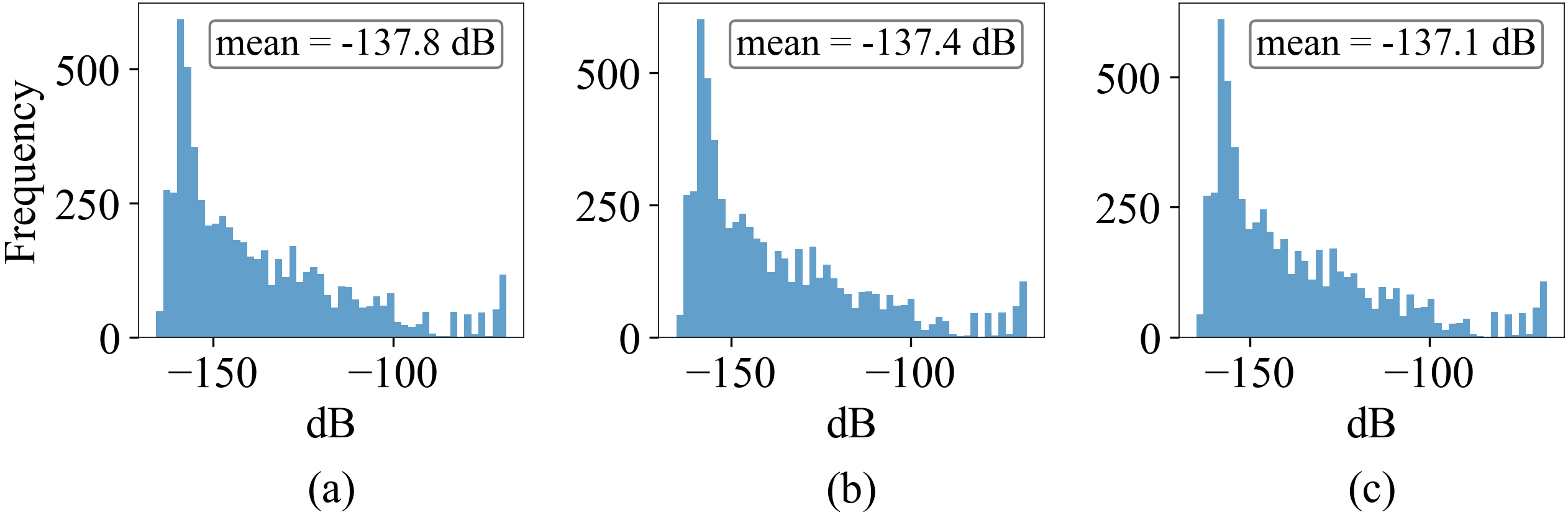}
  \caption{Empirical amplitude histograms (in dB) over all pixels for the three per-RX Doppler--time spectrograms. The means are similar but not identical, suggesting that each receive patch sees a slightly different multipath power distribution rather than an exact copy of the same signal.}
  \label{fig:spec3_hist}
\end{figure}

\subsection{Representation Modes and Classifier}
\label{subsec:repr_classifier}

Three input modes derived from the same radar measurements are compared. All inputs to the CNN are treated as tensors of shape $(C \times H \times W)$, where $C$ is the number of channels, $H$ the vertical dimension (Doppler or range), and $W$ the horizontal dimension.

\textbf{All-receiver spectrogram (\texttt{spec}):} For each clip, the digital twin exports an all-receiver Doppler--time spectrogram, which is .converted to decibels, cropped or zero-padded along the time axis to 26 frames, and z-scored per sample. The resulting tensor has shape $1 \times 26 \times 256$, where $H=26$ indexes slow-time frames (one frame per coherent processing interval, CPI) and $W=256$ indexes Doppler bins.

\textbf{Per-receiver spectrogram stack (\texttt{spec3}):} Three per-RX Doppler--time spectrograms (one per receive antenna) are loaded, converted to decibels, aligned, resized to $26 \times 256$, stacked, and z-scored per sample, yielding tensors of shape $3 \times 26 \times 256$. Here $H=26$ indexes slow-time frames, $W=256$ indexes Doppler bins, and $C=3$ encodes per-antenna diversity. The per-RX amplitude histograms in Fig.~\ref{fig:spec3_hist} differ slightly, consistent with three distinct receive patches.

\textbf{Static time-pooled range--Doppler map (\texttt{rdm}):} Let $X_t(r,q)$ denote the complex range--Doppler map at slow-time index $t$ for a given receive channel, where $r$ indexes range bins, $q$ indexes Doppler bins, and $t=1,\dots,T$ indexes all CPIs in a clip. A time-pooled magnitude map is defined as
\begin{equation}
  X_{\text{pool}}(r,q) = \max_{t=1,\dots,T} \bigl|X_t(r,q)\bigr|,
  \label{eq:rdm_timepool}
\end{equation}
Then $X_{\text{pool}}(r,q)$ is averaged over receive channels, converted to decibels, resized, and z-scored per sample.
The resulting tensors have shape $1 \times 256 \times 64$, where $H=256$ indexes Doppler bins and $W=64$ indexes range bins.
Temporal evolution within the clip is removed by the max over $t$, making \texttt{rdm} a static ``snapshot'' baseline.

Raw clips range from 22 to 53 slow-time frames (approximately $0.6$–$1.4$~s at $T_{\text{CPI}} = 26.3~\text{ms}$), depending on motion duration. To ensure uniform input dimensions, each clip is standardized to $T = 26$ frames by cropping longer sequences or zero-padding shorter ones, yielding a fixed temporal window of approximately $0.68$~s. This yields $2{,}880$ clips and $2{,}880 \times 26 = 74{,}880$ frames in total (about $0.55$~h of simulated motion).
\begin{table}[t]
  \centering
  \caption{Input dimensions and interpretation of the $H$ and $W$ axes for each radar mode (balanced dataset of $2{,}880$ clips).}
  \label{tab:repr_summary}
  \small
  \setlength{\tabcolsep}{3pt}
  \renewcommand{\arraystretch}{0.95}
  \begin{tabular}{@{}lccccc@{}}
    \toprule
    Mode & Shape $(C\!\times\!H\!\times\!W)$ & $H$ axis       & $W$ axis       & Clips \\
    \midrule
    \texttt{spec}   & $1\times 26\times 256$ & Time frames     & Doppler bins   & 2{,}880 \\
    \texttt{spec3}  & $3\times 26\times 256$ & Time frames     & Doppler bins   & 2{,}880 \\
    \texttt{rdm}    & $1\times 256\times 64$ & Doppler bins    & Range bins     & 2{,}880 \\
    \bottomrule
  \end{tabular}
\end{table}

\begin{figure}[t]
  \centering
  \includegraphics[width=\columnwidth]{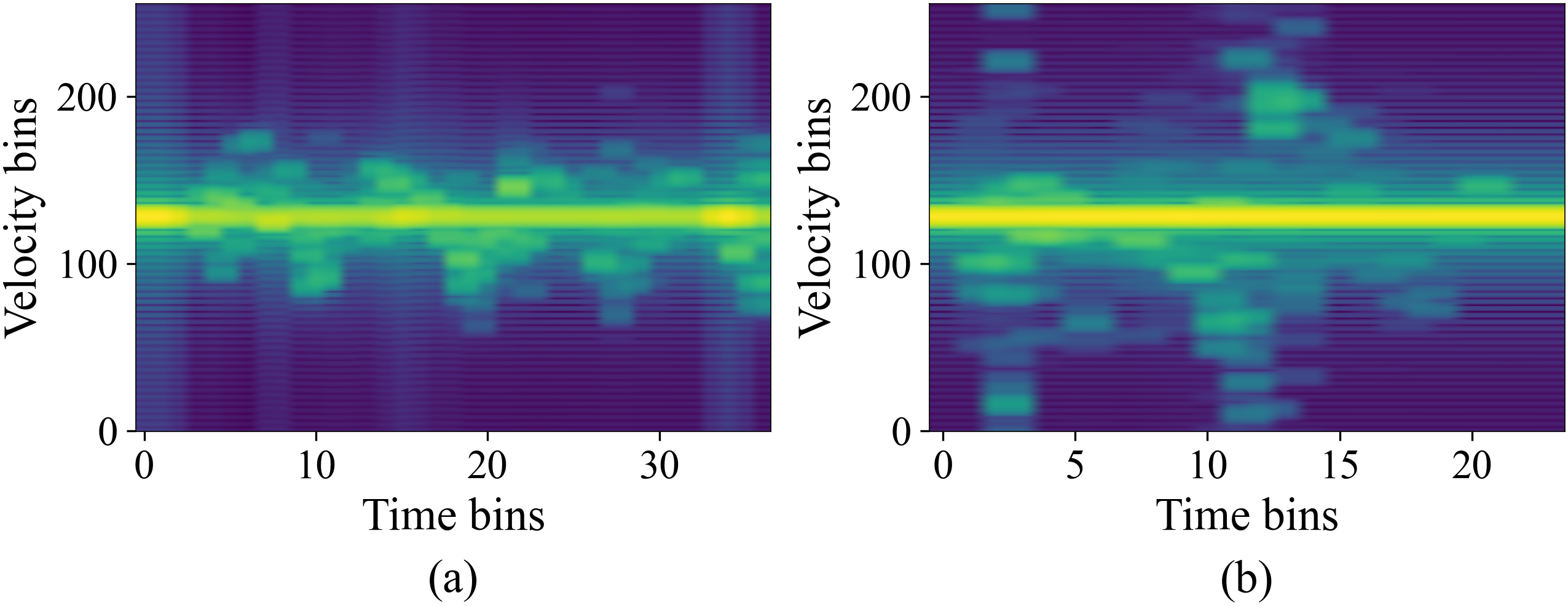}
  \caption{Example Doppler--time spectrograms at the fixed range bin $r^\star \approx 3.9~\text{m}$ for a non-fall (a) and a fall (b) clip in the digital twin.}
  \label{fig:spec_examples}
\end{figure}

Figure~\ref{fig:spec_examples} shows Doppler--time spectrograms at the fixed range bin for one non-fall and one fall, highlighting the stronger transient velocities during falls.

\subsubsection{CNN probe and training protocol}

All experiments use the same compact CNN, SmallFallCNN, with only the input channel count $C$ adapted to the mode. The network has three convolutional blocks (each: $3\times 3$ convolution, batch normalization, ReLU, and $2\times 2$ max-pooling with 16, 32, and 64 filters) followed by adaptive global average pooling to $(1,1)$ and a fully connected layer with two outputs (fall / non-fall). For \texttt{spec} and \texttt{rdm} ($C=1$) this yields 23{,}650 parameters; for \texttt{spec3} ($C=3$) it yields 23{,}938 parameters.

Models are implemented in PyTorch and trained from scratch with cross-entropy loss and the Adam optimizer (learning rate $10^{-3}$). Batch size is 16 for single-channel modes and 8 for \texttt{spec3}. Training runs for 100 epochs; after each epoch the model is evaluated on the fixed test split and at epoch 100 confusion matrices and per-class precision, recall, and F1-score are computed.

\section{Results}
\label{sec:results}

\begin{figure}[h]
  \centering
  \includegraphics[width=0.95\columnwidth]{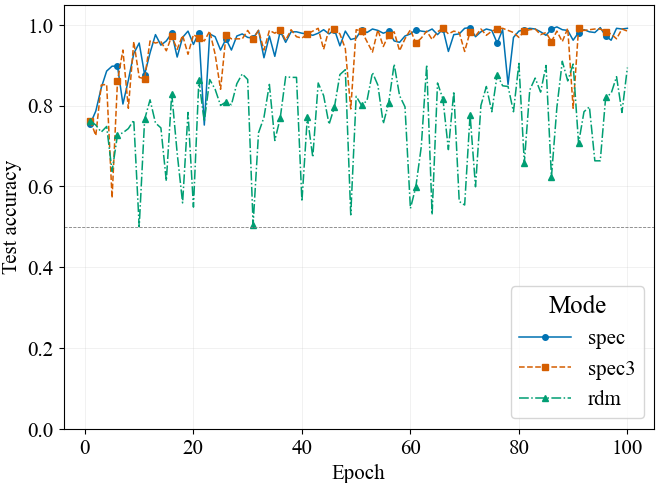}
  \caption{Test accuracy versus epoch for each representation mode. The spectrogram modes approach and stay near $99\%$, while the time-pooled RDM mode follows a lower curve with larger fluctuations over epochs.}
  \label{fig:train_curves}
\end{figure}

All results use the balanced 80/20 train/test split described in Section~\ref{sec:method}, with $N_{\text{train}} = 2{,}304$ clips and $N_{\text{test}} = 576$ clips (288 falls and 288 non-falls). For each mode, the corresponding SmallFallCNN is trained under the common protocol and evaluated on the test set after each epoch. Figure~\ref{fig:train_curves} shows the resulting test-accuracy curves. The spectrogram-based modes (\texttt{spec} and \texttt{spec3}) pass $95\%$ test accuracy within the first 10--20 epochs and then remain near $98$--$100\%$ for the rest of training, with only small fluctuations. The static RDM mode (\texttt{rdm}) also improves well above chance but exhibits much larger oscillations and settles noticeably below the spectrogram curves.

\begin{figure}[h]
  \centering
  \includegraphics[width=\columnwidth]{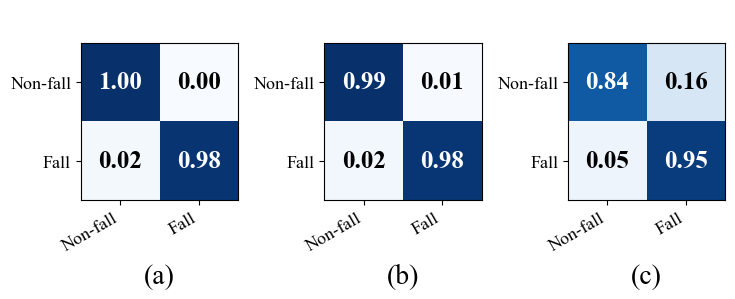}
  \caption{Normalized confusion matrices on the balanced test set for (a) \texttt{spec}, (b) \texttt{spec3}, and (c) \texttt{rdm} (rows: true labels, columns: predicted).}
  \label{fig:cm_all}
\end{figure}

Table~\ref{tab:perf} summarizes the final test performance after 100 epochs, including overall accuracy and class-wise precision and recall for non-fall (NF) and fall (F). All three modes perform reasonably well on this in-domain synthetic task, but the spectrogram modes are consistently stronger and more balanced across classes. The static RDM mode attains $89.4\%$ accuracy and noticeably lower non-fall recall, indicating more false alarms even though its overall accuracy remains high.

\begin{table}[t]
  \centering
  \caption{Test performance on the balanced synthetic test set ($N_{\text{test}}=576$) after 100 epochs.}
  \label{tab:perf}
  \small
  \setlength{\tabcolsep}{4pt}
  \renewcommand{\arraystretch}{0.9}
  \begin{tabular}{@{}lccccc@{}}
    \toprule
    Mode & Acc.\ (\%) & Prec.\ (NF) & Rec.\ (NF) & Prec.\ (F) & Rec.\ (F) \\
    \midrule
    \texttt{spec}   & 99.1 & 0.983 & 1.000 & 1.000 & 0.983 \\
    \texttt{spec3}  & 98.4 & 0.976 & 0.993 & 0.993 & 0.976 \\
    \texttt{rdm}    & 89.4 & 0.945 & 0.837 & 0.854 & 0.951 \\
    \bottomrule
  \end{tabular}
\end{table}

Figure~\ref{fig:cm_all} shows normalized confusion matrices for all three modes. For the spectrogram-based modes, diagonal entries are close to one and off-diagonal errors are rare. For the RDM mode, fall recall remains high, but non-fall recall drops, so the classifier generates more false alarms (non-falls misclassified as falls) despite a respectable overall accuracy. Under the fixed CNN probe and preprocessing used here, temporal spectrograms therefore provide more reliable fall/non-fall discrimination than the static RDM mode in this digital-twin setting.

\section{Discussion}
\label{sec:discussion}

In this digital twin, the gap between spectrogram and static RDM performance is consistent with the underlying motion. Falls produce a brief high-velocity descent followed by near-stationary returns, while non-fall activities such as walking, crouching, or sitting show slower, more distributed Doppler changes. Spectrogram modes preserve this Doppler--time evolution, so the CNN can exploit the shape and timing of the trajectory rather than only peak values.

The max-pooling in~\eqref{eq:rdm_timepool} removes this evolution before classification. A fast fall and a vigorous non-fall motion that reach similar peak velocities can generate similar pooled RDMs, encouraging the model to label many high-energy clips as falls and reducing non-fall recall. More advanced ways of combining information over time for RDMs (e.g., mean pooling, temporal windows, or learned attention) could alter this trade-off, but exploring such variants is outside the scope of this study. Note that the RDM max-pools over all available frames in each clip (22–53 depending on motion duration), while spectrograms are cropped to the first 26 frames; despite access to more temporal information, the static RDM still underperforms the spectrogram modes.

The three-channel spectrogram stack (\texttt{spec3}) does not substantially outperform the single-channel all-receiver spectrogram (\texttt{spec}) under the present conditions. With a single subject, high SNR, and a compact $1\times3$ array, the receive channels are strongly correlated and the combined spectrogram already captures most of the discriminative structure. Scenarios with multiple occupants, lower SNR, or larger arrays may benefit more from per-RX diversity.

% Spectrogram comparison
\begin{figure}[htb]
  \centering
  \includegraphics[width=\columnwidth]{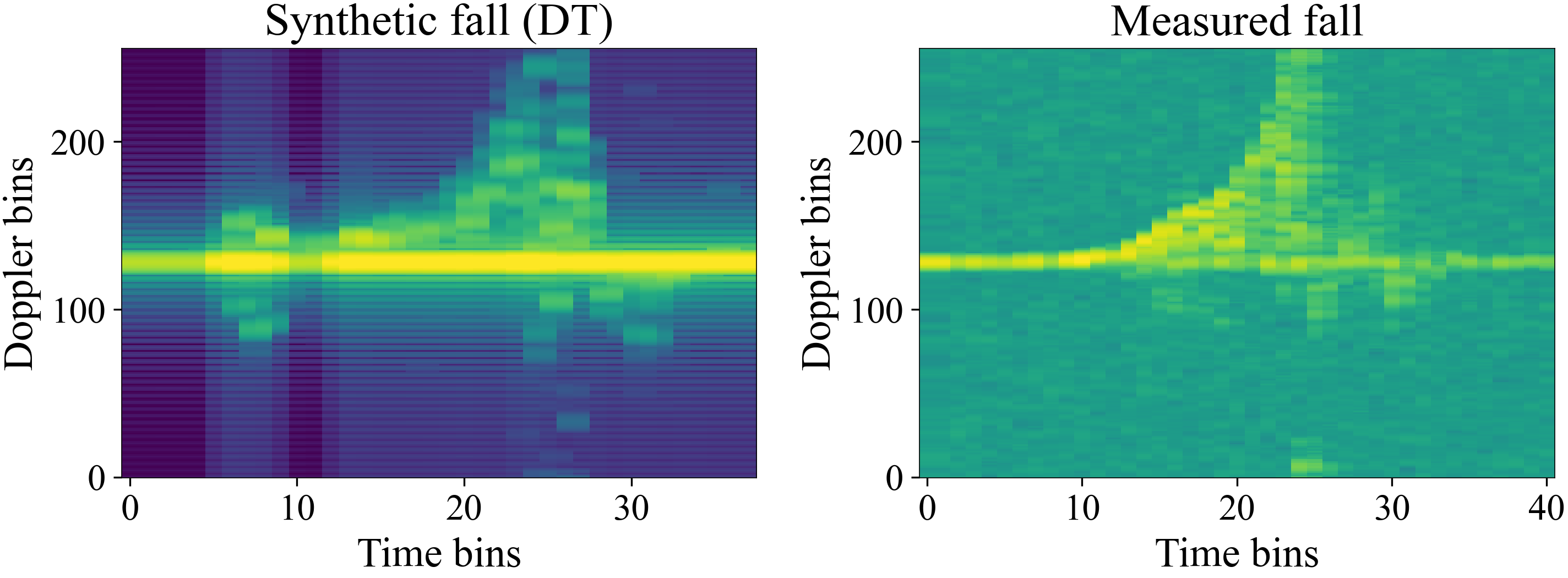}
  \caption{Doppler--time spectrograms for one fall: (left) synthetic from the digital twin and (right) measured from a 60~GHz radar. Both show a roughly triangular Doppler--time pattern during descent, followed by near-stationary returns; this is a qualitative check on gross signal structure.}
  \label{fig:real_sim_spec}
\end{figure}

% Histogram comparison
\begin{figure}[htbp]
  \centering
  \includegraphics[width=0.8\columnwidth]{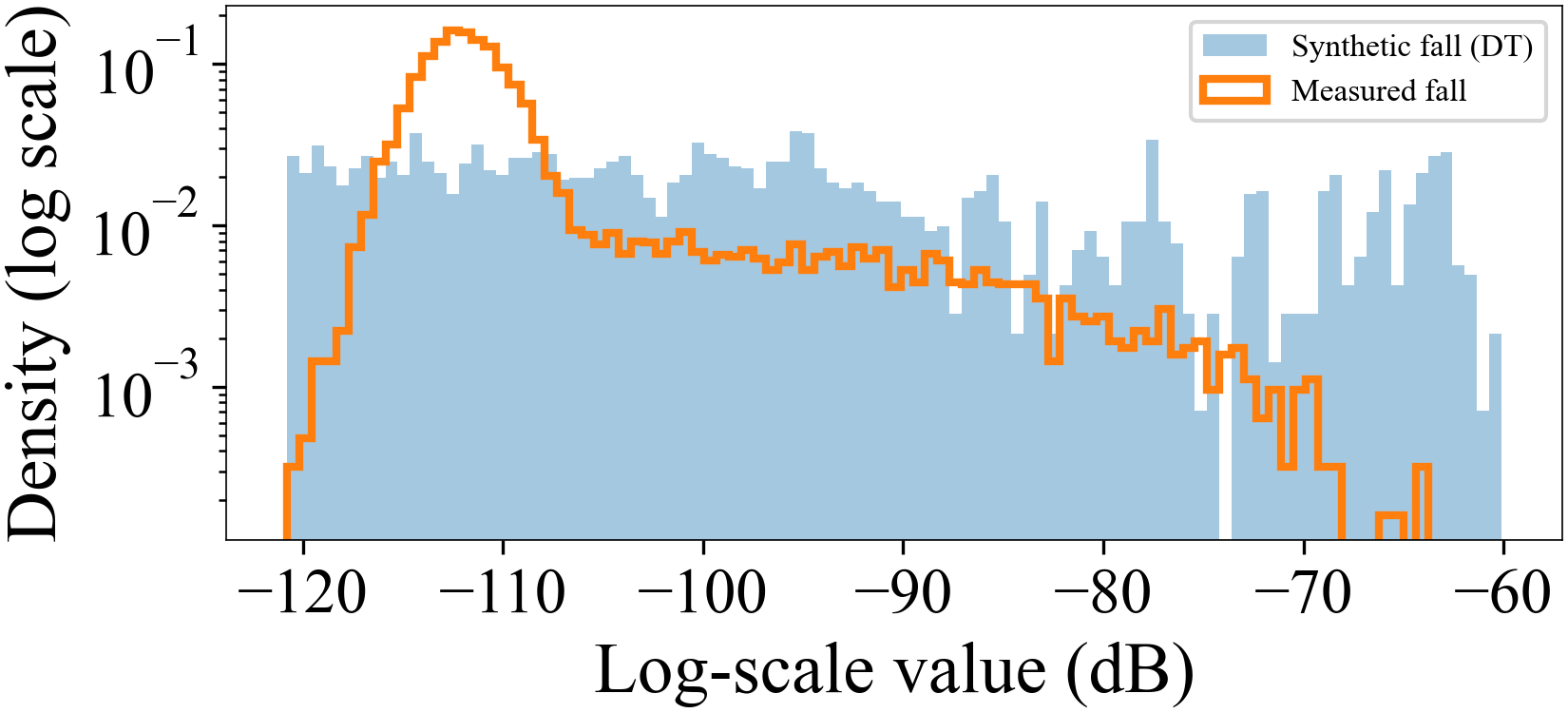}
  \caption{Pixel-wise amplitude histograms (log-scale density) for the synthetic and measured fall spectrograms in Fig.~\ref{fig:real_sim_spec}. For this example, the measured data show a tighter distribution around the modal value, consistent with receiver processing that the idealized twin omits.}
  \label{fig:real_sim_hist}
\end{figure}

To check whether the digital twin generates plausible fall signatures, Fig.~\ref{fig:real_sim_spec} compares one synthetic and one measured fall spectrogram, and Fig.~\ref{fig:real_sim_hist} compares their pixel-wise amplitude histograms. The Doppler--time patterns during descent are qualitatively similar, but the amplitude distributions differ, with the measured data showing a tighter cluster around a modal value. This is consistent with receiver processing (e.g., MTI filtering, AGC, quantization) and environmental factors that the idealized twin omits. The comparison is deliberately limited to a single measured example and is intended only as a plausibility check, not as a statistical validation of sim-to-real transfer.

Several limitations constrain how the results should be interpreted. The twin omits explicit receiver noise, hardware impairments, and non-stationary clutter; the motion library has more fall than non-fall templates and a narrow set of everyday activities; the dataset is artificially balanced, whereas real deployments are dominated by non-fall behaviour; only one room, radar placement, and waveform configuration are considered; and models are trained and tested within the same synthetic domain. Under these conditions, the near-99\% accuracies for spectrogram modes should be read as best-case, in-domain algorithmic behaviour in an idealized twin, not as clinical performance estimates. The main insight is the relative ranking of representations under controlled conditions, not the absolute risk of missed or spurious alarms in practice.

\section{Conclusion}
\label{sec:conclusion}

This paper presented a controlled digital-twin study of radar representations for fall detection in a single furnished room with a $60~\text{GHz}$ FMCW radar. A Perceive-EM workflow was used to simulate fall and non-fall motions across multiple positions and orientations, yielding a balanced synthetic dataset with 2{,}880 clips (1{,}440 falls and 1{,}440 non-falls). From the same slow-time data, three input modes were derived: an all-receiver Doppler--time spectrogram (\texttt{spec}), a per-receiver spectrogram stack (\texttt{spec3}), and a static range--Doppler map (\texttt{rdm}) obtained by max-pooling over time. Each mode was fed to an identical compact CNN probe and trained under the same optimization protocol on a fixed stratified train/test split.

In this setting, temporal spectrogram representations achieved the highest and most balanced test performance, with accuracies near $99\%$, while the static RDM representation reached $89.4\%$ accuracy and exhibited reduced non-fall recall and more volatile training curves. For the twin and motion library studied, Doppler--time dynamics appear more informative for fall discrimination than static range--Doppler snapshots aggregated over a few seconds of motion. A qualitative comparison between one synthetic and one measured fall spectrogram suggests that the twin captures gross Doppler--time structure, while histogram differences highlight a gap in amplitude statistics that is consistent with unmodeled receiver processing and environmental effects.

The study is restricted to an idealized single-room digital twin, a limited non-fall activity set, a balanced label distribution, and qualitative measured-data comparison only. Future work will extend the representation comparison to multi-room configurations and multiple radar placements, incorporate explicit noise and clutter models, broaden the non-fall motion library, and evaluate the same representations on measured datasets with ground-truth fall labels. Within these constraints and under tight model budgets, the results point toward temporal spectrograms as a strong baseline within this single-room digital-twin setting, while leaving open whether more advanced temporal aggregation for RDMs can close some of the observed gap.

\section*{Acknowledgment}

The authors would like to acknowledge CMC Microsystems, MITACS, Rogers, and the Ansys Electronics Desktop team for their support.

\bibliographystyle{IEEEtran}
\bibliography{clean_reference1}

\end{document}